\documentclass[aps,prd,twocolumn,groupedaddress,nofootinbib]{revtex4-2}

\usepackage{graphicx,graphics}
\usepackage{dcolumn}
\usepackage{bm}
\usepackage{xcolor}
\usepackage{soul}
\usepackage{hyperref}
\usepackage{amsmath}

\begin{document}


\title{Jovian Signal at BOREXINO}


\author{Saeed Ansarifard}
\email{ansarifard@ipm.ir}

\author{Yasaman Farzan}
\email{yasaman@theory.ipm.ac.ir}
\affiliation{School of physics, Institute for Research in Fundamental Sciences (IPM)
\\
P.O.Box 19395-5531, Tehran, Iran}


\date{\today}

\begin{abstract}
The BOREXINO experiment has been collecting solar neutrino data since 2007, providing the opportunity to study the variation of the event rate over a decade. We find that at 96 \% C.L., the rate of low energy events shows a time modulation  favoring a correlation with a flux from  Jupiter. We present a new physics model,  the  Jovian Whisper Model, based on dark matter of mass $\sim 0.1-4$~GeV captured by Jupiter that can account for such modulation. We discuss how the Jovian Whisper Model  (JWM) can be tested. 
\end{abstract}


\def\d{{\rm d}}
\def\Epos{E_{\rm pos}}
\def\ap{\approx}
\def\eff{{\rm eft}}
\def\L{{\cal L}}
\newcommand{\vev}[1]{\langle {#1}\rangle}
\newcommand{\CL}   {C.L.}
\newcommand{\dof}  {d.o.f.}
\newcommand{\eVq}  {\text{EA}^2}
\newcommand{\Sol}  {\textsc{sol}}
\newcommand{\SlKm} {\textsc{sol+kam}}
\newcommand{\Atm}  {\textsc{atm}}
\newcommand{\Chooz}{\textsc{chooz}}
\newcommand{\Dms}  {\Delta m^2_\Sol}
\newcommand{\Dma}  {\Delta m^2_\Atm}
\newcommand{\Dcq}  {\Delta\chi^2}
\newcommand{\nbb}{$\beta\beta_{0\nu}$ }
\newcommand {\be}{\begin{equation}}
\newcommand {\ee}{\end{equation}}
\newcommand {\ba}{\begin{eqnarray}}
\newcommand {\ea}{\end{eqnarray}}
\def\VEV#1{\left\langle #1\right\rangle}
\let\vev\VEV
\def\e6{E(6)}
\def\10{SO(10)}
\def\21{SA(2) $\otimes$ U(1) }
\def\321{$\mathrm{SU(3) \otimes SU(2) \otimes U(1)}$ }
\def\lr{SA(2)$_L \otimes$ SA(2)$_R \otimes$ U(1)}
\def\422{SA(4) $\otimes$ SA(2) $\otimes$ SA(2)}

\def\roughly#1{\mathrel{\raise.3ex\hbox{$#1$\kern-.75em
      \lower1ex\hbox{$\sim$}}}} \def\lsim{\roughly<}

\def\gsim{\roughly>}
\def\ltap{\raisebox{-.4ex}{\rlap{$\sim$}} \raisebox{.4ex}{$<$}}
\def\gtap{\raisebox{-.4ex}{\rlap{$\sim$}} \raisebox{.4ex}{$>$}}
\def\lsim{\raise0.3ex\hbox{$\;<$\kern-0.75em\raise-1.1ex\hbox{$\sim\;$}}}
\def\gsim{\raise0.3ex\hbox{$\;>$\kern-0.75em\raise-1.1ex\hbox{$\sim\;$}}}

\maketitle

\section{Introduction\label{intro|}}
Accumulation of BOREXINO data on solar $^7$Be neutrino line over a decade \cite{BOREXINO:2022wuy} provides the opportunity to study the time variation of the flux on Earth. Obviously, because of the eccentricity of the Earth orbit, $\epsilon = (1.6698 \pm 0.0003)\%$ \cite{1994A&A...282..663S}\footnote{https://data.giss.nasa.gov/modelE/ar5plots/srorbpar.html} the flux should have a $\sim 3.3\%$ annual variation. Although the fluctuations in the background of the experiment may contaminate this signal, thanks to the installation of the temperature control and stabilization system, the environmental conditions in the detector have been stabilized since 2019 \cite{BOREXINO:2020aww}. There are also slowly decreasing radioactive backgrounds which have been significantly reduced by the 2 final years of data taking \cite{BOREXINO:2020aww}.
 
Using the accurate BOREXINO data obtained after 2019, the annual variation of $^7$Be flux on Earth yields $\epsilon=(0.86\pm 0.62)\%$ 
which is about half the actual eccentricity. On the other hand, with the  data taken during 2011$-$2013, the result tends to be 1.6 times the actual value: $\epsilon=(2.74\pm 0.77)\%$.
This tension cannot be attributed to the fluctuations inside the Sun because the solar g-modes have much shorter periodicity and average to zero over one year \cite{Appourchaux:2009fe}. It is also worth noting that the contribution of the cosmogenic $^{11}C$ due to the cosmic ray is negligible in the  energy range under study \cite{BOREXINO:2022wuy}.

We entertain the possibility that the largest planet in the solar system, Jupiter, may play a role. Indeed, we find that associating an event rate of $\sim 1.5$ count per day per 100 tons of target $(\rm cpd/100t)$ to Jupiter, 
the observed change in the modulation of events over years can be explained.
Such extra contribution can be accommodated within the average flux uncertainty \cite{Vinyoles:2016djt}.
Measuring  the direction of the flux over time, this hypothesis can be tested. While dark matter with a mass smaller than 4~GeV can be trapped inside Jupiter, it will evaporate from the Sun \cite{Leane:2021tjj,Garani:2021feo,Li:2022wix,French:2022ccb,Blanco:2023qgi}. We build a model based on this Jovian feature to explain an event rate of few per day at BOREXINO associated to Jupiter. In our model, which we call Jovian Whisper Model (JWM), the annihilation of trapped DM pairs produces a flux of electromagnetically interacting particles that leads to a signal at BOREXINO mimicking the solar $^7$Be events.

In sect. \ref{sig}, we  discuss the modulation of the low energy BOREXINO event rate in the presence of a Jovian contribution and present the best fit for the intensity of this new component. In sect. \ref{sec:model}, we  introduce the Jovian Whisper Model that accounts for such contribution, respecting all the present bounds. In sect. \ref{Dis}, we  summarize the results and highlight approaches to test  the model. In appendix \ref{appen}, we demonstrate that a Jovian component significantly improves the fit to the solar neutrino data. In appendix \ref{accum}, we elaborate on the process of capture and the accumulation of light dark matter in Jupiter. We then discuss the impact of JWM in the early universe.
\section{Hint for a signal from Jupiter\label{sig}}
\begin{figure}
	\centering
	\includegraphics[width=0.45\textwidth]{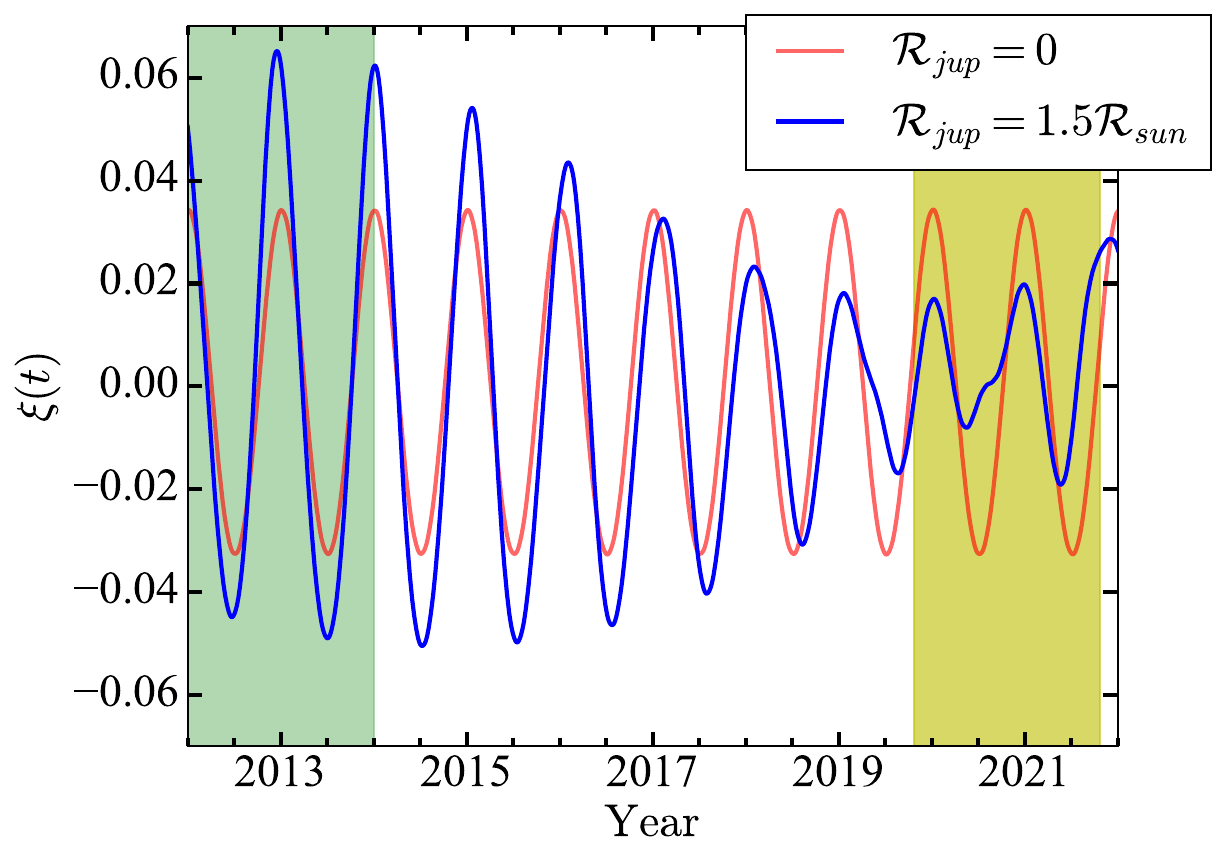}
		\caption{Relative modulation of the event rates  at the detector due to the orbital motion of the planets. The red (blue) curves shows the modulation without (with) a contribution from Jupiter. 
  The green area to the left corresponds to the period when Jupiter is on the side of the perihelion of the Earth orbit.
  The olive area to the right indicates the  period for which   Jupiter is located closer to the aphelion. During this period, the background conditions at the detector are most stable. }
		\label{fig:F1}
\end{figure}
In the presence of background and a contribution from Jupiter, the $\beta$-like event rate associated to the solar $^7$Be detection can be parameterized as 
\begin{equation}\label{eq:eventrate}
\mathcal{R}(t) = \frac{\mathcal{R}_{sun}}{d^2_{sun}(t)} +  \frac{\mathcal{R}_{jup}}{d^2_{jup}(t)} + \mathcal{R}_B
\end{equation}
where $d_{sun}$ ($d_{jup}$) is the distance from the Sun (Jupiter) to the Earth in Astronomical Unit (AU) as a function of  time $t$ which can be obtained using Ref. \cite{2019ascl.soft07024R}. $\mathcal{R}_{sun}$ and $\mathcal{R}_{jup}$ are respectively proportional to the event rates induced by Sun and Jupiter. We take them to be constant in time which is a reasonable assumption for the study of the annual variation. $\mathcal{R}_B$ accounts for the background contamination due to the radioactive sources. The main  sources  of the background in the energy range of interest comes from the decays of unstable isotopes  $\rm ^{210}Bi $  and $\rm ^{85}Kr$ \cite{BOREXINO:2022abl}. Analyzing the total (time-integrated) data,  the rates of the background from $\rm ^{210}Bi $  and $\rm ^{85}Kr$ are  extracted to be $4 \ (\rm cpd/100t)$ and  $3 \ (\rm cpd/100t)$, respectively \cite{BOREXINO:2022abl}. Nevertheless, as discussed before $\mathcal{R}_B$ at BOREXINO suffered a non-negligible variation in the time period before 2019.

The temporal average of $\mathcal{R}(t)$ over years can be approximately written as
\begin{equation}\label{eq:ave_flux}
\langle \mathcal{R} \rangle \approx \dfrac{\mathcal{R}_{sun}}{(1 \ {\rm AU})^2} + \dfrac{\mathcal{R}_{jup}}{(5 \ {\rm AU})^2} + \langle \mathcal{R}_B \rangle
\end{equation}
As long as $\mathcal{R}_{jup} \lsim \mathcal{ R}_{sun} $, the deviation of $\langle \mathcal{R} \rangle$ from the standard prediction will be within uncertainties of the prediction \cite{Gonzalez-Garcia:2023kva}. Neglecting the time variation of $\mathcal{R}_B$, we define a quantity demonstrating the modulation in the event rate due to varying $d_{sun}$ and $d_{jup}$
\begin{equation}
\xi(t) \equiv \dfrac{\mathcal{R}(t) - \langle \mathcal{R} \rangle}{[\mathcal{R}_{sun}/(1~{\rm AU})^2] }.
\end{equation}
We expect when Jupiter is located closer to Earth aphelion (around 2020), the annual variation of $\mathcal{R}(t)$ to be suppressed relative to the case that Jupiter is on the opposite side (around 2013) because in the latter case (when Jupiter is closer to perihelion), the annual maximums  of $\mathcal{R}_{sun}/d_{sun}^2$ and $\mathcal{R}_{jup}/d_{jup}^2$ add up. Fig.~(\ref{fig:F1}) shows the behavior of $\xi(t)$ consistent with this expectation. 
 
Utilizing the data from \cite{BOREXINO:2022wuy}, we concentrate on the period Oct 2019 to Oct 2021 in which the temporal variation of $\mathcal{R}_B$ is negligible \cite{BOREXINO:2020aww}, and try to constrain the free parameters in Eq.~(\ref{eq:eventrate}). Within the standard solar model and standard neutrino mass and mixing scheme, the total value of solar neutrinos in the region of interest is constrained  to ${\mathcal{R}_{sun}}/{(1 \ {\rm AU})^2} = 25\pm 2 \ (\rm cpd/100t)$ \cite{Esteban:2020cvm,Vinyoles:2016djt}. We implement this value as an input for our analysis with a Gaussian prior. We use flat prior $[0,50]$ for parameters $\mathcal{R}_{jup}$ and $\mathcal{R}_B$, excluding the constraint we have on $\mathcal{R}_B$ from the spectral fit. With this choice, a possible correlation between $\mathcal{R}_B$ and $\mathcal{R}_{jup}$ can be revealed.  Performing a Bayesian analysis \cite{2019ascl.soft10019T,Torrado:2020dgo}, we find the best fit value for $\mathcal{R}_{jup}$ as
\begin{equation}\label{eq:bestfit}
\frac{\mathcal{R}_{jup}}{(5 \ {\rm AU})^2} = 1.5^{+0.7}_{-0.8} \ (\rm cpd/100t) \ .  
\end{equation}

\begin{figure}
	\centering
	\includegraphics[width=0.4\textwidth]{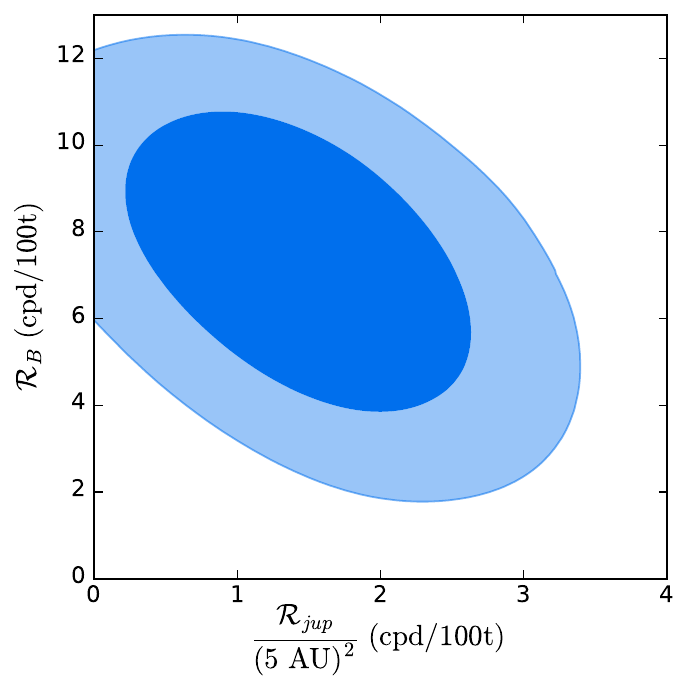}
		\caption{$68\%$ and $95\%$ constraint contour on the free parameters $\mathcal{R}_{jup}$ and $\mathcal{R}_B$. The analysis is performed using Eq.~(\ref{eq:eventrate}) and the last two years of data from \cite{BOREXINO:2022wuy}. The predicted value of $\mathcal{R}_{sun}$ is included with a Gaussian prior.}
		\label{fig:F2}
\end{figure}

The joint contour illustrated in Fig~(\ref{fig:F2}) shows an anti-correlation in the $\mathcal{R}_B- \mathcal{R}_{jup} $ plane. The central value of $\mathcal{R}_B \sim 7 \ \rm (cpd/100t) $ which is derived from spectral fit of the total (time-integrated) data \cite{BOREXINO:2023ygs} is consistent with our results based on time variation data. The Jovian flux can mimic the CNO events at BOREXINO so in the presence of the Jovian flux, the data would lead to a slightly smaller CNO flux fit which is in better agreement  with the solar model prediction(s) \cite{BOREXINO:2022abl}.  Furthermore the likelihood ratio test statistics independently disfavor the null Jovian signal ({\it i.e.,} $\mathcal{R}_{jup}=0$) at more than $96\% \ \rm C.L$ (see  appendix A for more detail).

\begin{figure}
	\centering
	\includegraphics[width=0.5\textwidth]{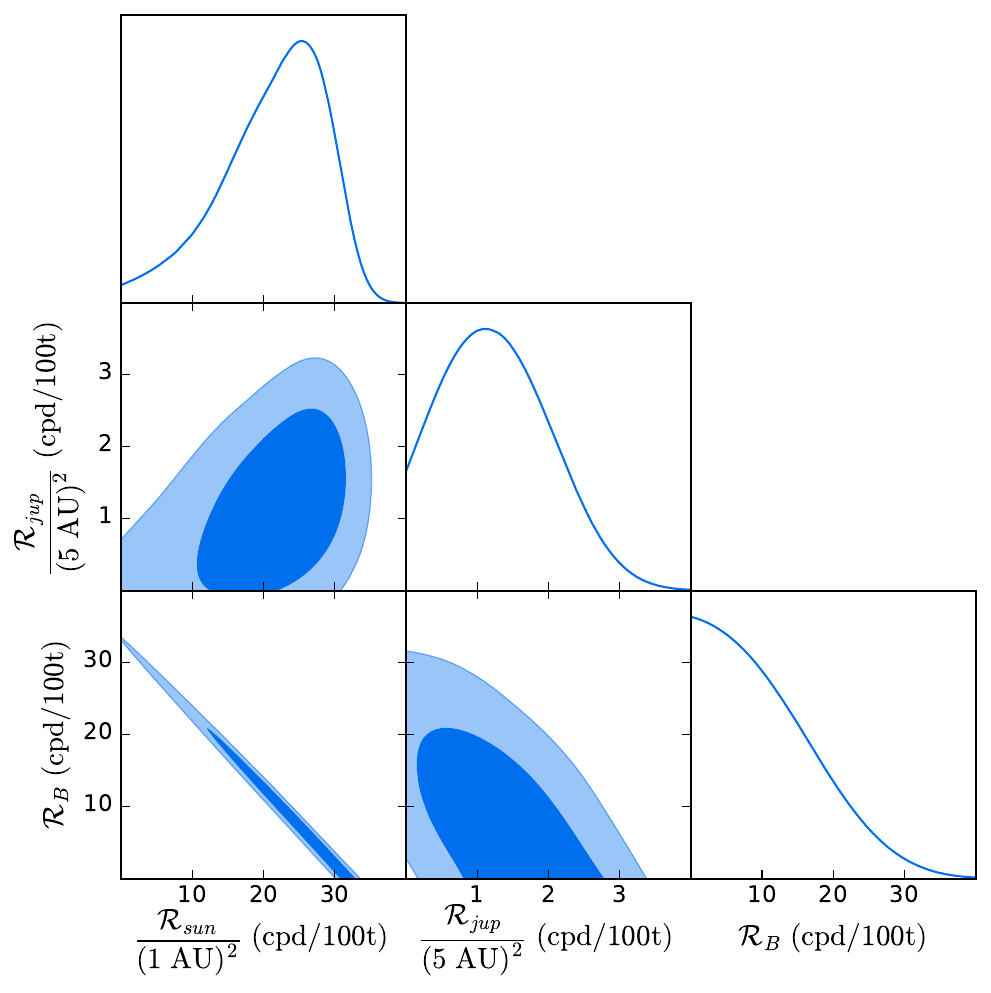}
		\caption{$68\%$ and $95\%$ constraint contour on the free parameters $\mathcal{R}_{jup}$, $\mathcal{R}_{sun}$ and $\mathcal{R}_B$. The analysis is performed using Eq.~(\ref{eq:eventrate}) and the last two years of data from \cite{BOREXINO:2022wuy}. The flat prior $[0,50]$ is considered for all parameters.}
		\label{fig:F3}
\end{figure}

We repeated the analysis of the time variation of the data during Oct 2019- Oct 2021 without a Gaussian prior on $\mathcal{R}_{sun}$ taking into account flat prior $[0,50]$. The results are shown in Fig ~(\ref{fig:F3}) which are consistent with the analysis carried out with a prior. Again the results indicate a non-zero value of $\mathcal{R}_{jup}$ at about $2\sigma$ C.L.

\begin{figure}
	\centering
	\includegraphics[width=0.5\textwidth]{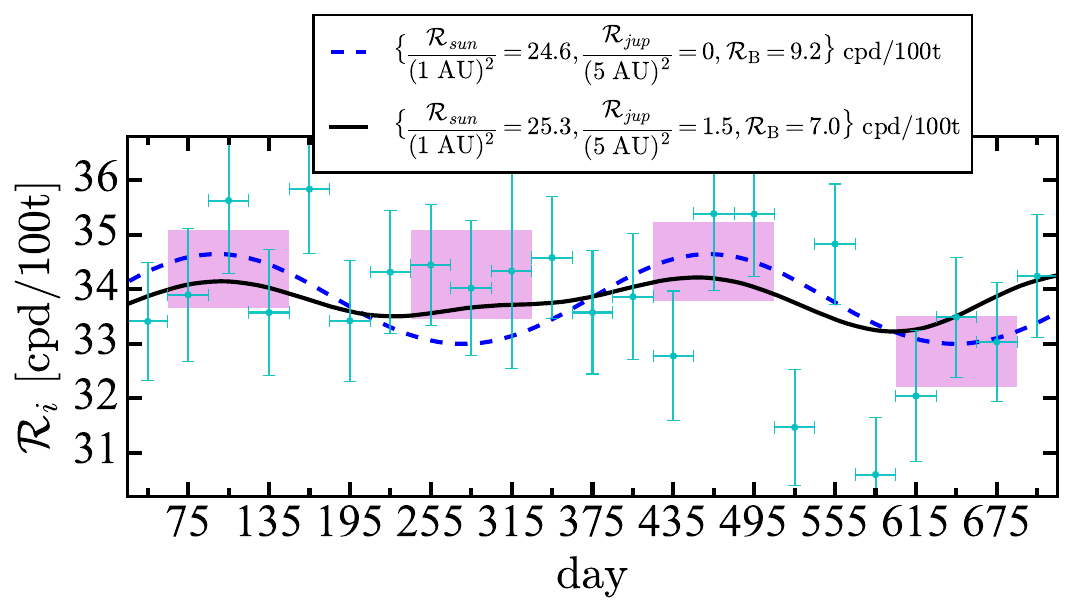}
		\caption{Monthly-binned $\beta$-like event rate in the energy range corresponding to the $\rm ^7Be$ solar neutrinos from Oct 2019 to Oct 2021 obtained by the BOREXINO collaboration \cite{BOREXINO:2022wuy}. The black line (blue dash-line) shows the prediction with (without)  the presence of the  Jovian Signal.}
		\label{fig:monthly}
\end{figure}

Monthly-binned data are shown in Fig~(\ref{fig:monthly}). The solid black  and  dashed blue curves respectively show the predictions  for the time variation with and without the Jovian contribution. The highlighted boxes are three-months averaged data around aphelion and perihelion.

\begin{figure}
	\centering
	\includegraphics[width=0.4\textwidth]{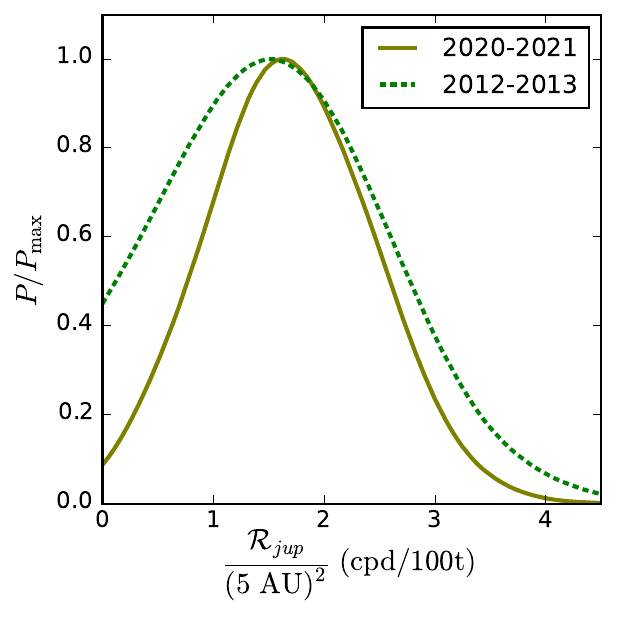}
		\caption{The marginalized posteriors on $\mathcal{R}_{jup}$ in period Oct 2019 to Oct 2021 (olive line) and Dec 2011 to Dec 2013 (green dash-line) using modulation monthly binning data \cite{BOREXINO:2022wuy}.}
		\label{fig:F5}
\end{figure}

So far we have concentrated on the time interval with a negligible background variation ({\it i.e.,} 2019-2021). During the whole period of the BOREXINO data taking from 2007, the time variation of  $\mathcal{R}_B$ has been non-negligible. However, the shape of time dependence of $\mathcal{R}_B$ is known. Following the approach introduced by the collaboration, we decompose the event rate into a trend and modulation as follows.
\begin{equation}
\mathcal{R}(t) = R_{tr}(t) + \delta \mathcal{R}(t)
\end{equation}
where $R_{tr}(t)$ is the trend of the data with a monotonic behavior known up to some parameters that are determined by  fitting  to the data. After removing the trend, we use the modified version of Eq.~(\ref{eq:eventrate}) to probe the annual modulation
\begin{equation}\label{eq:eventrate_trend}
\delta \mathcal{R}(t) = \sum_i  \mathcal{R}_{i} \left[\frac{1}{d^2_{i}(t)} - \int_{\rm exp} \frac{dt}{d^2_{i}(t)}\right]
\end{equation}
where $i$ runs over $sun$ and $jup$ and the integral is taken over the  time period of interest. The free parameters are taken to be $\mathcal{R}_{jup}$ and  $\mathcal{R}_{sun}$,  with the same prior as before. We conduct a similar Bayesian analysis to constrain $\mathcal{R}_{jup}$ using the modulation data which is provided by the BOREXINO collaboration \cite{BOREXINO:2022wuy}. We neglect the effect of the $R_{tr}(t)$ on the result. This is a reasonable assumption as long as the background does not have a periodic variation. This approach is somehow similar to  removing the average of the data for the study of time variation for the  Oct 2019 to Oct 2021 period when the background was constant. However,  the trend is computed using the whole period of the experiment. From Fig \ref{fig:F1}, we observe that during the period 2015-2018, the Jovian contribution is not expected to alter the modulation. The deviation due to the Jovian contribution is however non-negligible  during periods 2011-2013 and 2019-2021. We therefore focus on these two periods to extract  $\mathcal{R}_{jup}$. As seen in Fig. \ref{fig:F1},  we expect to see an enhancement instead of damping in the modulation during 2011-2013. In this period the trend quite differs from the average due to  the fast decay of some of the radioactive contamination. After marginalizing over $\mathcal{R}_{sun}$, the normalized posterior of $\mathcal{R}_{jup}$ is computed separately for these two period. The results  are shown in Fig~(\ref{fig:F5}). Interestingly, the two years of the data, taken during the expected  enhancement of the annual modulation (2011-2013), also favors the solution  found by analyzing the data from 2019-2021 by $\sim 2\sigma$. Best fit value for  the time modulation of the data taken during the last two years (Oct 2019 to Oct 2021)  is
\begin{equation}
\frac{\mathcal{R}_{jup}}{(5 \ {\rm AU})^2} = 1.7^{+0.8}_{-0.8} \ (\rm cpd/100t) \ . \label{last}
\end{equation}
Similarly for the first two years from Dec 2011 to Dec 2013, we obtain
\begin{equation}
\frac{\mathcal{R}_{jup}}{(5 \ {\rm AU})^2} = 1.6^{+0.8}_{-1.1} \ (\rm cpd/100t) \ .\label{first}
\end{equation}
The values in Eqs. (\ref{last},\ref{first}) are in remarkable agreement.
Similar analysis is performed on SNO annual variation data which spans the period Nov 1999 to Oct 2003 and the energy range $\sim 5-20 \ \rm MeV$  \cite{SNO:2005ftm}. In case of the presence of a Jovian signal in the $\rm ^8B$ energy range, this period coincides with the enhancement of  the annual modulation. Considering the constraints on $\mathcal{R}_{sun}$ and $\mathcal{R}_B$ from the SNO spectral fit \cite{SNO:2005oxr} as priors and assuming a stable background, we have found an upper limit on the Jovian signal in the $5-20$ MeV  range as
\begin{equation}
\frac{\mathcal{R}_{jup}}{(5 \ {\rm AU})^2} < 1.2 \ (\rm cpd/1000t) \ at \ 99\% C.L.\ .
\end{equation} 
We should emphasize that the SNO data differs from that of BOREXINO  in at least two aspects. First, while  the BOREXINO data is composed of  $\beta$-like elastic scintillator events, the SNO data comes from the Cherenkov radiation which is sourced by the charged current and neutral current interactions on the deuterium as well as elastic scattering events of the electrons in the detector. Second, the energy ranges of these two experiments are different. To avoid the stringent bound from SNO, it should be ensured that the   Jovian Whisper Model (JWM) does not lead to the electron recoil energy larger than 5~MeV which is the detection threshold of SNO.

\section{Jovian Whisper Model (JWM) \label{sec:model}}
In this section, we propose a toy model based on the absorption of Dark Matter (DM) particles ($\chi$) in Jupiter and their subsequent annihilation to intermediate particles that give rise to a signal at BOREXINO, mimicking that of the solar $^7$Be neutrinos. If $\chi$ with  a mass of $\sim 1$~GeV  or lighter composes less than 1 $\%$ of DM ({\it i.e.,} $f_\chi<1 \%$)  the spin-dependent  cross section of $\chi$ scattering off the protons, $\sigma_p$ can be as large as $\sim 10^{-24}$ cm$^2$ \cite{Li:2022idr} without violating the recently found bounds \cite{Ema:2024oce,Roy:2024ear}.
The mean free path of $\chi$ at the surface of Jupiter can be written as
\begin{equation} 
	\left( (\rho_s^J/m_p)[(1+f_H)\sigma_p/2+(1-f_H)\sigma_n/2]\right)^{-1} \label{mean}
	\end{equation}
where $\rho^J_s$ is the surface density of Jupiter and $f_H$ is its Hydrogen fraction. $\sigma_p$  and $\sigma_n$ are respectively the  cross sections of scattering of $\chi$ off protons and neutrons.
With a matter density of  $\rho_s^J=$1.3 ${\rm gr/cm^3}$ and being $f_H=$71 \%  composed of Hydrogen, the mean free path of $\chi$ with $\sigma_p\sim 10^{-29}$ cm$^2$ on the Jovian surface will be  2~cm. If the mean free path is much shorter than the Jovian radius,  all  $\chi$ particles reaching the Jupiter surface will be absorbed, yielding an absorption rate of $$\mathcal{A}=4\pi R_J^2 n_\chi v_\chi$$ where $R_J =70000$ km is the Jupiter radius,  $n_\chi=f_\chi (0.4~{\rm GeV/cm^3})/m_\chi$ and $v_\chi \sim 10^{-3} c$ so $$\mathcal{A}=7\times 10^{25} ~{\rm sec}^{-1} (f_\chi/0.01)({\rm GeV}/m_\chi).$$
In appendix \ref{accum}, we discuss the bounds on $\sigma_n$ and $\sigma_p$ in the mass range of interest and present a model for $\sigma_n \sim 10^{-29}$ cm$^2$.  With such cross section, the mean free path at the Jupiter surface will be less than 10~km which means that  not only $\chi$ can  be trapped in the Jupiter surface but can be even thermalized. We then discuss that as long as $m_\chi$ is heavier than about 100 MeV, the velocity of the majority of the thermalized $\chi$ will be smaller than the escape velocity so they will be accumulated by Jupiter. We show that incorporating the $17$ MeV solution to ATOMKI anomaly \cite{Alves:2023ree} yields $\sigma_n\sim 10^{-29}$ cm$^2$.

The accumulated $\chi$ may annihilate to hadrons through the same coupling responsible for scattering off nuclei. However, if we assume asymmetric scenario where $n_{\chi}\gg n_{\bar{\chi}}$, this mode of annihilation will be closed. Instead, we couple $\chi$ to a  scalar $\phi$ with a mass $m_\phi \sim $few 10 MeV through the following scalar potential
\begin{equation}  V=m_\phi^2 |\phi|^2+(\lambda_{\phi \chi} |\phi|^2 \chi^2 +H.c.) \  .
\end{equation}
 Notice that the  global $U(1)$ symmetry under which   $\chi \to \exp (-i \alpha ) \chi$ is broken by the quadratic $\lambda_{\phi \chi}$ coupling but the $Z_2$ symmetry ($\chi \to -\chi$) remains unbroken and protects $\chi$ from decay.
 The $\lambda_{\phi \chi}$ coupling can lead to $\chi \chi \to \phi \bar\phi$. For $\lambda_{\phi\chi}<1$,  $$\sigma_\chi=\sigma(\chi \chi \to \phi \bar{\phi}) <3 \times 10^{-29} \ {\rm cm}^2.$$ In order for $\chi \chi \to \phi \bar\phi$ to balance the $\chi$ capture in Jupiter, the following relation should hold 
 $$ \mathcal{A}=2 v_\chi^J (n_\chi^J)^2 V_J\sigma_\chi \  , $$
where $V_J=4\pi R_J^3/3$. $v_\chi^J$ and $n_{\chi}^J$ are respectively the velocity and density of $\chi$ inside Jupiter. Taking the age of Jupiter equal to $t_J=4.5\times 10^9$ years, we can write $n_\chi^J\leq \mathcal{A} t_J/V_J$.  In the limit that all captured $\chi$ particles remain inside Jupiter the equality holds valid. The 
velocity of $\chi$ should be  around the thermal velocity in the Jupiter center ($\sim 10~{\rm km/sec}$).
Putting these together, we find $\sigma_\chi \geq 2\times 10^{-37} {\rm cm}^2 (0.01/f_\chi)(0.1~{\rm GeV}/m_\chi)^{1/2}$. 

A short discussion of the $\phi$ and $\chi$ production in the early universe is given in Appendix \ref{accum}. Before discussing how the $\phi$ particles can lead to a signal in BOREXINO, let us check whether the $\chi$ capture in the celestial bodies can dramatically warm their surface. The rate of energy absorption due to the $\chi$ capture per unit area is $\rho_\chi v_\chi$ which is the solar system is $10^{-5} {\rm GeV/(cm^2 sec)}$. This value should be compared to    the black body radiation rate from the surface of Jupiter ($\sigma_{SB}T^4=2.5\times 10^7{\rm GeV/(cm^2 sec)}$). Neutron stars have typically much hotter surface so the energy absorption due to the $\chi$ capture will be much smaller than its black body radiation. Even if all $\chi$ energy was converted to kinetic energy (which is not the case in our model), it would not drastically change the surface temperature of the celestial bodies.


The $\phi$ pairs from annihilation of $\chi$ pairs will have an energy  equal to $m_\chi$. The source of the Jovian signal detected by BOREXINO cannot be the direct scattering of $\phi$ coming from Jupiter at BOREXINO. If this was the case, a much larger flux of $\phi$ originating from the $\chi$ capture in the Earth should have been detected. In the end of this section we introduce two alternative scenarios which are applicable in two different ranges of the $\phi$ and $\chi$ masses, both explaining the Jovian signal at BOREXINO. In both scenarios, each $\phi$ particle produced inside Jupiter leads to production of $\mathcal{N}$ pairs of magntic dipole particle $C\bar{C}$ with a mass of $\sim 10$ MeV.  The $C$ particles with mass of $\sim 10$ MeV and magnetic dipole of  $\mu_C\sim 3~ {\rm TeV}^{-1}$ are still allowed by existing bounds \cite{Chang:2019xva}. The differential cross section of a $C$ particle with an energy of $E_C$ colliding on an electron can be written as \begin{equation}  \frac{d\sigma_e}{dT}=\frac{\alpha \mu_C^2}{8k^2}\frac{m_C}{\sqrt{k^2+m_e^2}}\frac{2k^2+Tm_e}{T},\end{equation}
where  $T$ is  the recoil energy, $k$ is the momentum in the center of mass frame, $k=(s-m_C^2)/(2\sqrt{s})$ and $s=m_e^2+m_C^2+2m_eE_C$. 
The scattering of $C$ off electrons inside BOREXINO with a recoil energy less than $\sim 1$~MeV  ($T \lsim 1$~MeV) can mimic solar $^7$Be neutrino events.  The cross section of this process for $m_e^2\ll m_C^2$ can be estimated as 
$\sigma_e\simeq 5 \alpha \mu_C^2$ \cite{Chang:2019xva}. The mean free path of $C$ particles with $\mu_C=3/$TeV is larger than the diameter of the Earth so they will reach the detector. A fraction of $C$ coming from the Jupiter may scatter off protons inside the Earth which will change their direction but not their energy.

The rate of electron scattering events per day per 100 tonnes  with a recoil energy less than $\sim$1 MeV is expected to be
\begin{equation}\label{rate}
\mathcal{F}_C \times {\rm day} \times \frac{\sigma_e}{4\pi d_{jup}^2}\times \frac{100~{\rm t}}{2m_p}.
\end{equation} 
where $\mathcal{F}_C$ is the sum of the  fluxes of $C$ and $\bar{C}$ particles: 
\begin{equation}
\mathcal{F}_C=2\mathcal{A} \mathcal{N},\label{FC}
\end{equation}	
where $\mathcal{N}$ is the average number of $C\bar{C}$ pairs at BOREXINO, originating from a single $\phi$ produced in Jupiter. In what follows, we introduce two alternative scenarios for the production of $C$ from $\phi$ and discuss the value of $\mathcal{N}$ for each scenario. We also discuss how a large flux of $C$ from the $\chi$ capture by Earth can be avoided.

\paragraph{Scenario I---}
{\it Model with $m_\chi \sim 0.1$ GeV and $m_\chi -m_\phi< 20$ MeV:} 
In this scenario, $\phi$ with energy  equal to $m_\chi$ leaves Jupiter undisturbed and on its way to Earth decays  to a $\phi^\prime$ pair which in turn decay to $C\bar{C}$ pairs: $\phi \to \phi^\prime  \bar{\phi}^\prime \to C\bar{C}C\bar{C}$. The average energy of the final $C$ and $\bar{C}$ will be about  $m_\chi/4\sim 25$ MeV so the typical recoil energy will be around the recoil energy from $^7$Be neutrino interaction. Thus, the flux of $C\bar{C}$ can mimic the $^7$Be neutrino flux.  Taking the decay lengths $\tau_\phi$ and $\tau_{\phi^\prime}$ about 1 AU,  the majority of $\phi $ and $\phi^\prime$ particles decay {\it en route} to Earth, leading to $\mathcal{N}=4$.  The Jovian signal at BOREXINO per day will be
$$ \frac{\mathcal{R}_{jup}}{(5~{\rm AU})^2}=2 (\frac{f_\chi}{10^{-3}})\left(\frac{\mu_C}{3~{\rm TeV}^{-1}}\right)^2	 ~~ ({\rm cpd/100 t})
$$ 
which is enough for accounting for the observed hint.  The signal from  the $\chi$ capture in Earth will be suppressed by $(R_E/R_J)^2 (d_{jup}^2/R_E^2) [R_E^2 /(\tau_\phi\tau_{\phi^\prime})]=0.2$. Notice that we have taken $\tau_\phi\sim \tau_{\phi^\prime} \sim d_{jup}/5$. The flux from the Earth will be therefore too small to be resolved with present uncertainties.

\begin{figure}
	\centering
	\includegraphics[width=0.5\textwidth]{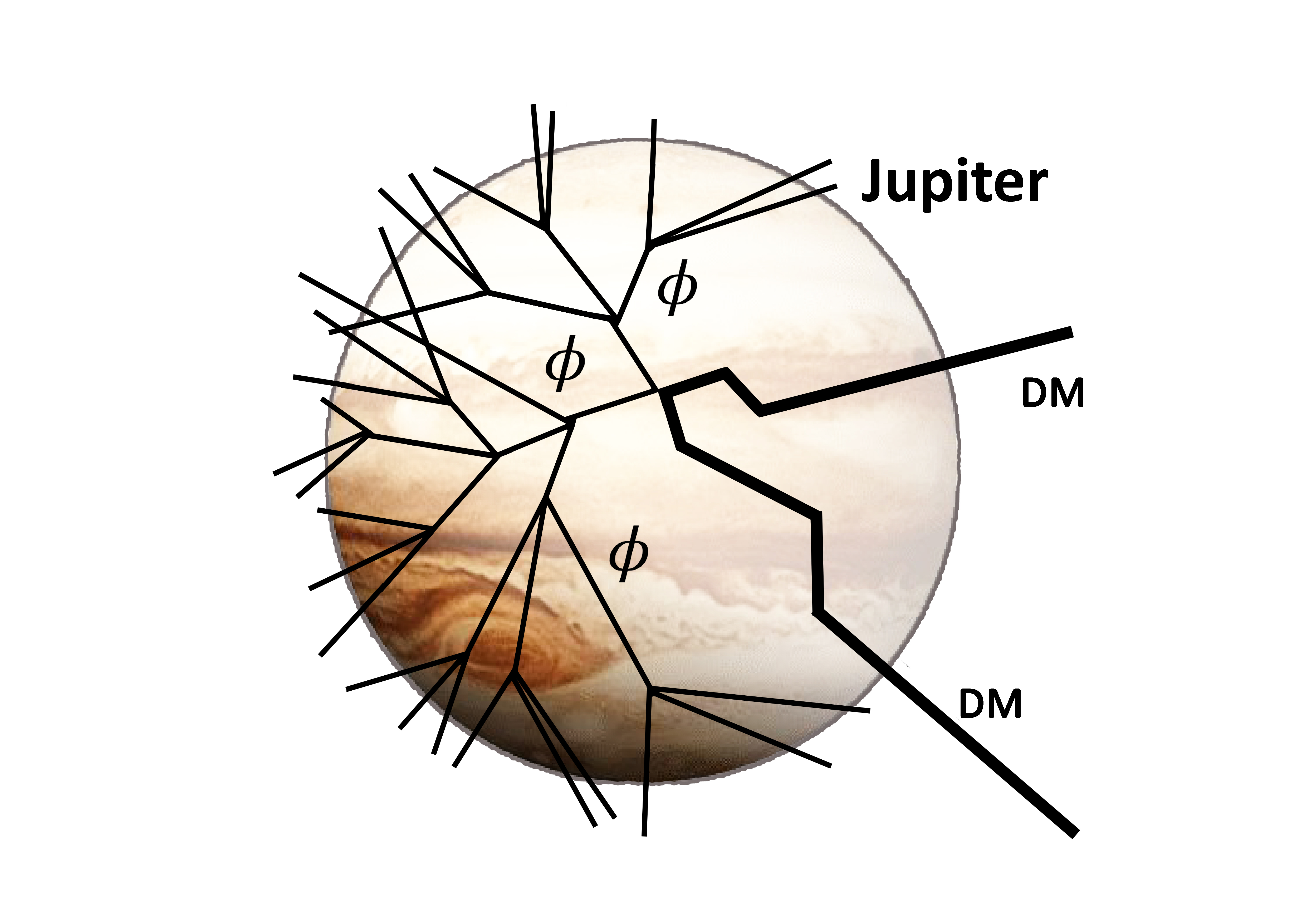}
		\caption{Schematic view of DM trapping inside Jupiter, its annihilation to mediators and their cascading down to lower energies as takes place in scenario II. }
		\label{fig:Jup}
\end{figure}

\paragraph{Scenario II---}
{\it Model with $m_\chi \sim$ GeV and $m_\phi\sim 20$ MeV:}
Let us suppose there is a new vector boson, $V$ with a mass of $(2-3)\times m_\phi$ and with  couplings of order of $g_\phi\sim O(1)$ to $\phi$ and of $g_n=O(10^{-3})$ to the  nucleons \cite{Gninenko:1998pm,NOMAD:1998pxi}.  Then, for  $m_\phi\sim m_V\ll \sqrt{s}\sim m_N$,
$$\sigma (\phi+N\to \phi +N)\sim \frac{g_\phi^2g_n^2}{2\pi m_{V}^2} \frac{\sqrt{s}}{m_N} \sim 10^{-32}  {\rm cm}^2  $$  and
\begin{equation}
   \dfrac{\sigma (\phi+N\to \phi +N+V)}{ \sigma (\phi+N\to \phi +N)}  \sim \frac{ g_\phi^2}{ \pi^2}  \left( \log \frac{\Delta E}{m_V}\right)^2\sim {\rm few} \times 10^{-2},
\end{equation}
where $\Delta E$ is the energy transfer to $N$.
With $g_\phi\sim 1$ and $ g_n\sim 10^{-3}$, $\phi$ can go through a few ($\sim 5$) bremsstrahlung scatterings before leaving Jupiter.
While in the two body scattering, the average energy loss is $O(m_V)$, in the bremsstrahlung scattering the kinetic energy of $\phi$ will be shared between $V$ and final $\phi$ and on average, the energy of $\phi$ can be reduced
to half.  $V$ will immediately decay into a $\phi \bar{\phi}$ pair so a shower of $\phi$ and $\bar{\phi}$ will be formed after $\sim 5$ bremsstrahlung scattering.  The energy of $\phi$ while leaving the planet will be therefore about $2^{-5} m_\chi \sim m_{V}$. When the kinetic  energy of $\phi$ drops below $m_V$, the bremsstrahlung process stops. Moreover, in these energies scattering off nuclei will be elastic and will not lead to a significant energy loss.
The $\phi$ produced via $\chi$ annihilation will then cascade down to $\sim m_\chi/m_V$ lower energy $\phi$ particles leaving Jupiter. We therefore expect a flux of  $\mathcal{A} (m_\chi/m_V) $  of $\phi$ particles with energies of $O(m_V)$ streaming out Jupiter. Fig. \ref{fig:Jup} schematically depicts the above processes.

In principle, the $\phi$ particles can electromagnetically interact with the electrons inside the detector but then we expect a large signal induced by $\phi$ coming from the Earth, too. To avoid such a contribution from the Earth, we take $\phi$ to be neutral but assume that it decays to a pair of dipole particles, $\phi \to C\bar{C}$ with a decay length, $\tau_\phi$ of order of 1 AU. Majority of the $\phi$ particles traveling from Jupiter at a distance of 5 AU will decay to $C\bar{C}$ pairs so $\langle\mathcal{N}\rangle= m_\chi/m_V$. Using Eqs.  (\ref{rate},\ref{FC}), the Jovian signal rate can therefore be estimated  as
\begin{equation}
	\frac{\mathcal{R}_{jup}}{(5 \ \rm AU)^2} \sim 2\times \frac{f_\chi}{ 10^{-3}}\left(\frac{\mu_C}{ 3~{\rm TeV}^{-1}} \right)^2 \ \rm cpd/100t \ .
\end{equation}
Since $\mathcal{A}$ is proportional to $m_\chi^{-1}$, the rate turns out to be independent of $m_\chi$.

Let us now estimate the rate of events induced by $\chi$ annihilation in Earth. The rate of $\chi$ capture and therefore the $\phi$ production in Earth is suppressed by $R_E^2/R_J^2$ relative to production in Jupiter. The $\phi$ particles do not lose considerable energy before leaving the Earth so they will not cascade to multiple $\phi$ particles inside the Earth. Moreover, their decay length will be boosted relative to the decay length of $\phi$ from Jupiter: $\sim \tau_\phi m_\chi/m_V$. The ratio of the number of events induced by $\phi$  from the Earth to that from Jupiter can be estimated as $$ \frac{R_E^2}{R_J^2}\frac{d_{jup}^2}{R_E^2} \left(\frac{m_V}{m_\chi}\right)^2 \frac{R_E}{\tau_C}$$ which for $\tau_C\sim 1$ AU  and $m_\chi>2$~GeV is smaller than 1 and therefore safe.





Unless stated otherwise, the following discussion applies  for both scenario I and II.
The $C$ particles can in principle lead to an electron excess at the direct dark matter search experiments such as XENONnT \cite{XENON:2022ltv}. Taking $\mathcal{R}_{jup}/(5~{\rm AU})^2=1.5$ cpd/100 t, we expect a rate of $3({\rm keV}/T)$ events/ton.year.keV,  which can be accommodated within the present uncertainties but can be tested by the improvements of the uncertainty by a factor of $O(3)$. If we took $C$ particles to be millicharged ({\it i.e.,} if we took monopole coupling of $C$ with photon rather than dipole coupling), we could still explain the Jovian signal but at lower recoil energies relevant for XENONnT, the number of recoiled electron events would exceed the observation. That is because while for the millicharged particles $d\sigma_e/dT \propto 1/T^2$, for the dipole particles, $d\sigma_e/dT \propto 1/T$. If the interaction of the $C $ particles with the electron is via a heavy mediator, $d\sigma_e/dT$ would have a milder dependence on $T$ so we would not expect a detectable signal at direct dark matter search experiments.

Let us now discuss how the bound from SNO can be avoided in our model. First remembering that the interaction of $C$ with matter fields takes place via a virtual photon exchange and the photon has  only vector (monopole non-axial) coupling to Deuterium, it cannot contribute to the Gamow-Teller Deuterium dissociation process. That is the number of neutral current events at SNO will not be affected. Moreover, with proper choice of parameters, the recoil energy of the electron in  elastic scattering can be lower than the energy threshold of SNO (5~MeV). In the scenario I, the maximum $E_C$ is $m_\chi -3m_C$. Taking $m_\chi=100$~MeV and $17~{\rm MeV}<m_C<20$~MeV, the recoil energy will be smaller than 5~MeV and the majority of the recoiled electrons will have an energy smaller than 0.8~MeV, mimicking the $^7$Be signal at BOREXINO. In scenario II, if we take $E_C<25$~MeV and $m_C\sim 10$~MeV, we obtain similar results.


\section{Discussion and conclusions\label{Dis}}
We have analyzed the BOREXINO $^7$Be neutrino data collected from 2011 to 2021 and found that the time variation of the signal deviates from the expected $1/d_{sun}^2$ behavior. In particular, the last time window of the data, whose background was well under control, yields a value for the eccentricity of the Earth orbit which is about half the known value. New physics scenarios with 1 year periodicity such as pseudo-Dirac neutrinos \cite{Ansarifard:2022kvy,Anamiati:2017rxw} are not able to resolve this tension.

To explain the unexpected behavior, we have hypothesized  a flux of new particles from Jupiter that can induce a signal similar to that of the solar $^7$Be flux with a $1/d_{jup}^2$ time variation. We found that with such contribution, the time variation of the signal at BOREXINO during 2019-2021 will be suppressed, explaining the lower extracted eccentricity. The best fit for the average rate of events from Jupiter is about 6\% of the $^7$Be events so it can hide in the 6\% uncertainty of the solar neutrino prediction. This hypothesis can be tested by measuring the direction of incoming flux(es) over time. Since the recoil energy of the electrons will be below the detection threshold of SNO (5~MeV),   the bound from SNO can be avoided.
Super-Kamiokande further suppresses the Jovian signal by selecting only signals from the Sun direction \cite{Super-Kamiokande:2023jbt,Super-Kamiokande:2023yqq}.

We have proposed a model that accounts for such signal from Jupiter. Our so-called Jovian Whisper Model (JWM)  is based on  testable ingredients: (1) A fraction of $O(10^{-3})$ of dark matter with mass of smaller than 4 GeV and with a rather strong scattering off nuclei which can be tested by future dark matter search experiments \cite{Li:2022idr}. Such dark matter component can be trapped by Jupiter but not by Sun \cite{Li:2022wix,French:2022ccb,buckley2015lhapdf6}. 
We have shown that the same 17 MeV boson that accounts for the ATOMKI anomaly can be responsible for the scattering of $\chi$ off nuclei.
 (2) The decay of intermediate particles from the annihilation of the trapped dark matter on its way to the Earth leads to a flux of dipole particles with a mass of $O(10)$ MeV and $\mu_C\sim 3~{\rm TeV}^{-1}$ (testable by ILC). The  flux of dipole particle leads to a signal at BOREXINO   mimicking the solar $^7$Be line. Since  the differential cross section is proportional to the inverse of recoil energy, upcoming electron excess measurements by direct dark matter search experiments can test this model. We have shown that a flux of dipole particles from dark matter capture in the Earth can be avoided if the decay length of the intermediate particles are of order of 1 AU. Our model introduces new neutral particles with masses of $\sim 10$ MeV. We have discussed how they can evade the bounds from cosmology on light new particles.

\begin{acknowledgments}
This work has been supported by the European Union$'$s Framework Programme for Research and Innovation Horizon 2020 under grant H2020-MSCA-ITN-2019/860881-HIDDeN as well as under the Marie Sklodowska-Curie Staff Exchange  grant agreement No 101086085-ASYMMETRY. The authors  would like to acknowledge support from ICTP through the Associates Programme and from the Simons Foundation through grant number 284558FY19. SA thanks A. M. Serenelli, J. Salvado and J.P. Pinheiro for the helpful discussion. Fig.~\ref{fig:F2}, Fig.~\ref{fig:F3} and Fig.~\ref{fig:F5} are produced using GetDist package \cite{Lewis:2019xzd}. The authors thank S. Jam for helping them to produce Fig.~\ref{fig:Jup}. The  authors are grateful to R. Leane and P. Ullio for useful comments.
\end{acknowledgments}

\appendix


\section{Hypothesis test statistics \label{appen}}
\begin{figure}
	\centering
	\includegraphics[width=0.45\textwidth]{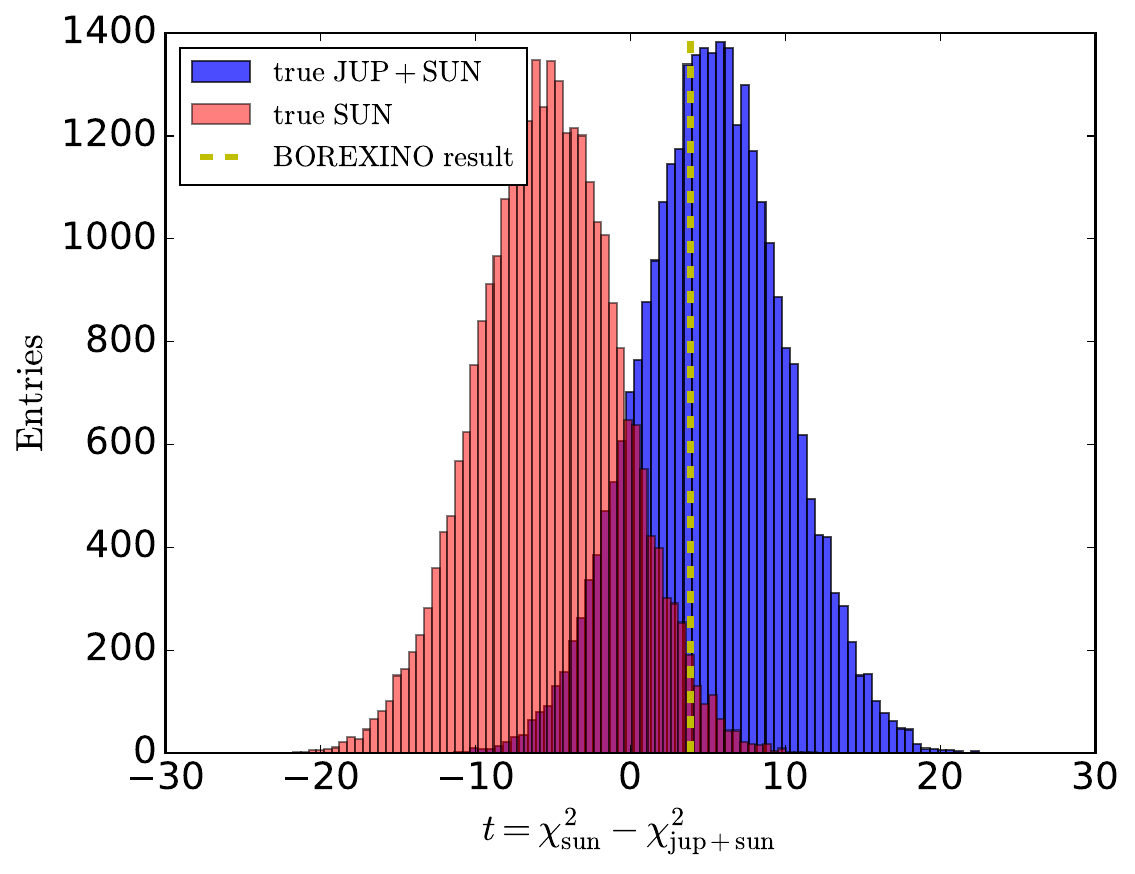}
		\caption{The distribution of the $\boldsymbol{t}$ statistics considering the true signal comes  only from Sun (red) or from both Sun and Jupiter (blue). The dashed line shows the BOREXINO result.}
		\label{fig:F6}
\end{figure}

In order to test whether the variation of data is because of the background uncertainties, we randomly generate 30000 sample of simulated monthly binned  data with  the Jovian signal, taking the best fit values for the parameters $$\{\frac{\mathcal{R}_{sun}}{(1 \ \rm AU)^2} = 25.3, \frac{\mathcal{R}_{jup}}{(5 \ \rm AU)^2} = 1.5 , \mathcal{R}_B = 7.0 \} \ (\rm cpd/100t)$$ and compare them with the best fit achieved by the standard case, taking $$\{\frac{\mathcal{R}_{sun}}{(1 \ \rm AU)^2} = 24.6, \mathcal{R}_{jup}=0, \mathcal{R}_B = 9.2\} \ (\rm cpd/100t).$$
Fig~\ref{fig:F6} shows the distribution of the $\boldsymbol{t}$ statistics which is defined as
\begin{equation}
\boldsymbol{t} = -2\log \left( \dfrac{\mathcal{L}({\rm SUN})}{\mathcal{L}({\rm SUN+JUP})}\right)
\end{equation}
where $\mathcal{L}(F)$ is the likelihood of theory $F$ being true. The BOREXINO result gives $\boldsymbol{t} =3.33$ which  disfavors the standard hypothesis with No-Jovian signal  at $96.5\%$ C.L. 

Unsurprisingly, the rest of  the BOREXINO data taken during 2015 to 2019 is not able to discriminate the Jovian signal due to the large uncertainties in the temporal variation of the background and high correlation of the signal with the solar neutrino seasonal variation. The data before 2010 highly suffers from backgrounds which makes them unsuitable for this analysis \cite{BOREXINO:2022wuy}.

\section{Implications of JWM for early universe cosmology and dark matter capture by Jupiter \label{accum}}
If the scattering cross section of dark matter is larger than certain values, it cannot reach the direct dark matter search detectors located deep underground.  As a result, for $f_\chi=0.01$ and $m_\chi\sim 1$ GeV, there is a window around spin dependent $\sigma_p\sim 10^{-24}$ cm$^2$ which is not still ruled out \cite{Li:2022idr,Ema:2024oce,Roy:2024ear}. For lighter $\chi$, there is no bound from direct dark matter search experiments but there are bounds from Cosmic Ray (CR) boosted argument \cite{Bringmann:2018cvk,Cappiello:2019qsw,Wang:2021nbf,PandaX-II:2021kai,CDEX:2022fig}. Taking $f_\chi=0.01$, $\sigma_p<10^{-30}$ cm$^2$ is allowed \cite{Li:2022idr} for $m_\chi=0.1$~GeV. Considering that $\sim 95 \%$ of CR (with energy of order of 10 GeV) is composed of protons, leaving  only 5 $\%$ for neutrons, we can conclude that $\sigma_n<{\rm few}\times 10^{-29}$~cm$^2$ is still allowed for $m_\chi \sim 0.1$~GeV.  With such a value of $\sigma_n$, the mean free path on the Jupiter surface will be  about 8 km$\ll R_J$ which implies that the $\chi$ particles can reach thermal (kinetic)  equilibrium with matter in the Jupiter surface: $m_\chi \langle (v_\chi^J)^2\rangle/2=3kT/2$. Let us define $ v_{tail}^2=100kT/ m_\chi$.  The fraction of $\chi$ inside Jupiter with $v>v_{tail}$ is of order of $10^{-21}$ and therefore completely negligible. Taking even the Jovian surface temperature $T=163$ Kelvin, we find $v_{tail}=35~{\rm km/sec} (100~{\rm MeV}/m_\chi)^{1/2}<v_{esc}=60~{\rm km/sec}$. This means that Jupiter can efficiently accumulate $\chi$ as light as 0.1~GeV provided that the mean free path is much shorter than 0.1 $R_J$. Since the surface of the Sun is much hotter, it cannot hold such light $\chi$.

The famous ATOMKI anomaly has a canonic solution which involves a new boson, $X$, with a mass of $m_X=17$~MeV coupled to both neutrons, $g_n$ and electrons, $g_e$ \cite{Alves:2023ree}. Allowing a coupling of $g_\chi$ between the $\chi$ and $X$ particles, we can write \begin{equation} \sigma_n \sim \frac{g_n^2g_\chi^2}{4\pi}\frac{m_\chi^2}{m_X^4}\sim 10^{-29}~{\rm cm}^2 g_\chi^2\left(\frac{g_n}{10^{-3}}\right)^2 \left(\frac{m_\chi}{100~{\rm MeV}}\right)^2
 \end{equation}

In the early universe, the $\chi \bar{\chi}$ pair can be produced and thermalized via the same interaction that captures $\chi$ in the Jupiter.  The  $\lambda_{\phi \chi}$ coupling  will then bring $\phi$ and $\bar{\phi}$ particles to thermal equilibrium with the plasma in the early universe when the temperature was above the masses of these particles.  Turning on a coupling between $\phi$ and $X$ ($g_\phi$) can lead to $\sigma(\phi \bar{\phi} \to e^-e^+)\sim g_e^2g_{\phi}^2/(4\pi (2 m_\phi)^2)\sim 10^{-35}~{\rm cm}^2 (g_e/10^{-4})^2 g_\phi^2 (80~{\rm MeV}/m_\phi)^2$ where $g_e$ is the coupling of $X$ to $e^-e^+$ with a value set equal to that in the solution to the ATOMKI anomaly \cite{Alves:2023ree}. However, the coupling of $\phi$ to $X$ may lead to its scattering inside Jupiter. To avoid such scatterings, the $\phi$ pairs may annihilate to $\nu \bar{\nu}$ pairs (instead of $e^-e^+$) in the early universe via coupling to a light mediator. If $\phi$ particles were stable after freeze-out, they would contribute 10 $\%$ to the DM energy budget but $\phi$ particles decay with a lifetime of $\sim 8$ minutes (after BBN) to particles that are dark and cannot dissociate nuclei. As a result, they safely avoid the bounds from CMB and BBN. The $\chi \bar{\chi}$ pairs in the early universe can also efficiently annihilate to $e^-e^+$ or $\phi \bar{\phi}$ via a coupling of form $|\phi|^2|\chi|^2$.  For $T>m_\chi$, the $\chi\chi \to \phi \bar{\phi}$ process can wash out the $\chi$ excess over  $\bar{\chi}$. As a result, the $\chi$ excess should be produced at $T<m_\chi$ with some out of equilibrium process  which is indeed one of Sakharov's conditions.

\bibliography{submit_v4.bib}

\begin{thebibliography}{38}%
\makeatletter
\providecommand \@ifxundefined [1]{%
 \@ifx{#1\undefined}
}%
\providecommand \@ifnum [1]{%
 \ifnum #1\expandafter \@firstoftwo
 \else \expandafter \@secondoftwo
 \fi
}%
\providecommand \@ifx [1]{%
 \ifx #1\expandafter \@firstoftwo
 \else \expandafter \@secondoftwo
 \fi
}%
\providecommand \natexlab [1]{#1}%
\providecommand \enquote  [1]{``#1''}%
\providecommand \bibnamefont  [1]{#1}%
\providecommand \bibfnamefont [1]{#1}%
\providecommand \citenamefont [1]{#1}%
\providecommand \href@noop [0]{\@secondoftwo}%
\providecommand \href [0]{\begingroup \@sanitize@url \@href}%
\providecommand \@href[1]{\@@startlink{#1}\@@href}%
\providecommand \@@href[1]{\endgroup#1\@@endlink}%
\providecommand \@sanitize@url [0]{\catcode `\\12\catcode `\$12\catcode
  `\&12\catcode `\#12\catcode `\^12\catcode `\_12\catcode `\%12\relax}%
\providecommand \@@startlink[1]{}%
\providecommand \@@endlink[0]{}%
\providecommand \url  [0]{\begingroup\@sanitize@url \@url }%
\providecommand \@url [1]{\endgroup\@href {#1}{\urlprefix }}%
\providecommand \urlprefix  [0]{URL }%
\providecommand \Eprint [0]{\href }%
\providecommand \doibase [0]{https://doi.org/}%
\providecommand \selectlanguage [0]{\@gobble}%
\providecommand \bibinfo  [0]{\@secondoftwo}%
\providecommand \bibfield  [0]{\@secondoftwo}%
\providecommand \translation [1]{[#1]}%
\providecommand \BibitemOpen [0]{}%
\providecommand \bibitemStop [0]{}%
\providecommand \bibitemNoStop [0]{.\EOS\space}%
\providecommand \EOS [0]{\spacefactor3000\relax}%
\providecommand \BibitemShut  [1]{\csname bibitem#1\endcsname}%
\let\auto@bib@innerbib\@empty
\bibitem [{\citenamefont {Appel}\ \emph {et~al.}(2023)\citenamefont {Appel}
  \emph {et~al.}}]{BOREXINO:2022wuy}%
  \BibitemOpen
  \bibfield  {author} {\bibinfo {author} {\bibfnamefont {S.}~\bibnamefont
  {Appel}} \emph {et~al.} (\bibinfo {collaboration} {BOREXINO}),\ }\href
  {https://doi.org/10.1016/j.astropartphys.2022.102778} {\bibfield  {journal}
  {\bibinfo  {journal} {Astropart. Phys.}\ }\textbf {\bibinfo {volume} {145}},\
  \bibinfo {pages} {102778} (\bibinfo {year} {2023})},\ \Eprint
  {https://arxiv.org/abs/2204.07029} {arXiv:2204.07029 [hep-ex]} \BibitemShut
  {NoStop}%
\bibitem [{\citenamefont {{Simon}}\ \emph {et~al.}(1994)\citenamefont
  {{Simon}}, \citenamefont {{Bretagnon}}, \citenamefont {{Chapront}},
  \citenamefont {{Chapront-Touze}}, \citenamefont {{Francou}},\ and\
  \citenamefont {{Laskar}}}]{1994A&A...282..663S}%
  \BibitemOpen
  \bibfield  {author} {\bibinfo {author} {\bibfnamefont {J.~L.}\ \bibnamefont
  {{Simon}}}, \bibinfo {author} {\bibfnamefont {P.}~\bibnamefont
  {{Bretagnon}}}, \bibinfo {author} {\bibfnamefont {J.}~\bibnamefont
  {{Chapront}}}, \bibinfo {author} {\bibfnamefont {M.}~\bibnamefont
  {{Chapront-Touze}}}, \bibinfo {author} {\bibfnamefont {G.}~\bibnamefont
  {{Francou}}},\ and\ \bibinfo {author} {\bibfnamefont {J.}~\bibnamefont
  {{Laskar}}},\ }\href@noop {} {\bibfield  {journal} {\bibinfo  {journal}
  {AAP}\ }\textbf {\bibinfo {volume} {282}},\ \bibinfo {pages} {663} (\bibinfo
  {year} {1994})}\BibitemShut {NoStop}%
\bibitem [{\citenamefont {Agostini}\ \emph {et~al.}(2020)\citenamefont
  {Agostini} \emph {et~al.}}]{BOREXINO:2020aww}%
  \BibitemOpen
  \bibfield  {author} {\bibinfo {author} {\bibfnamefont {M.}~\bibnamefont
  {Agostini}} \emph {et~al.} (\bibinfo {collaboration} {BOREXINO}),\ }\href
  {https://doi.org/10.1038/s41586-020-2934-0} {\bibfield  {journal} {\bibinfo
  {journal} {Nature}\ }\textbf {\bibinfo {volume} {587}},\ \bibinfo {pages}
  {577} (\bibinfo {year} {2020})},\ \Eprint {https://arxiv.org/abs/2006.15115}
  {arXiv:2006.15115 [hep-ex]} \BibitemShut {NoStop}%
\bibitem [{\citenamefont {Appourchaux}\ \emph {et~al.}(2010)\citenamefont
  {Appourchaux} \emph {et~al.}}]{Appourchaux:2009fe}%
  \BibitemOpen
  \bibfield  {author} {\bibinfo {author} {\bibfnamefont {T.}~\bibnamefont
  {Appourchaux}} \emph {et~al.},\ }\href
  {https://doi.org/10.1007/s00159-009-0027-z} {\bibfield  {journal} {\bibinfo
  {journal} {Astron. Astrophys. Rev.}\ }\textbf {\bibinfo {volume} {18}},\
  \bibinfo {pages} {197} (\bibinfo {year} {2010})},\ \Eprint
  {https://arxiv.org/abs/0910.0848} {arXiv:0910.0848 [astro-ph.SR]}
  \BibitemShut {NoStop}%
\bibitem [{\citenamefont {Vinyoles}\ \emph {et~al.}(2017)\citenamefont
  {Vinyoles}, \citenamefont {Serenelli}, \citenamefont {Villante},
  \citenamefont {Basu}, \citenamefont {Bergstr\"om}, \citenamefont
  {Gonzalez-Garcia}, \citenamefont {Maltoni}, \citenamefont {Pe\~na Garay},\
  and\ \citenamefont {Song}}]{Vinyoles:2016djt}%
  \BibitemOpen
  \bibfield  {author} {\bibinfo {author} {\bibfnamefont {N.}~\bibnamefont
  {Vinyoles}}, \bibinfo {author} {\bibfnamefont {A.~M.}\ \bibnamefont
  {Serenelli}}, \bibinfo {author} {\bibfnamefont {F.~L.}\ \bibnamefont
  {Villante}}, \bibinfo {author} {\bibfnamefont {S.}~\bibnamefont {Basu}},
  \bibinfo {author} {\bibfnamefont {J.}~\bibnamefont {Bergstr\"om}}, \bibinfo
  {author} {\bibfnamefont {M.~C.}\ \bibnamefont {Gonzalez-Garcia}}, \bibinfo
  {author} {\bibfnamefont {M.}~\bibnamefont {Maltoni}}, \bibinfo {author}
  {\bibfnamefont {C.}~\bibnamefont {Pe\~na Garay}},\ and\ \bibinfo {author}
  {\bibfnamefont {N.}~\bibnamefont {Song}},\ }\href
  {https://doi.org/10.3847/1538-4357/835/2/202} {\bibfield  {journal} {\bibinfo
   {journal} {Astrophys. J.}\ }\textbf {\bibinfo {volume} {835}},\ \bibinfo
  {pages} {202} (\bibinfo {year} {2017})},\ \Eprint
  {https://arxiv.org/abs/1611.09867} {arXiv:1611.09867 [astro-ph.SR]}
  \BibitemShut {NoStop}%
\bibitem [{\citenamefont {Leane}\ and\ \citenamefont
  {Linden}(2023)}]{Leane:2021tjj}%
  \BibitemOpen
  \bibfield  {author} {\bibinfo {author} {\bibfnamefont {R.~K.}\ \bibnamefont
  {Leane}}\ and\ \bibinfo {author} {\bibfnamefont {T.}~\bibnamefont {Linden}},\
  }\href {https://doi.org/10.1103/PhysRevLett.131.071001} {\bibfield  {journal}
  {\bibinfo  {journal} {Phys. Rev. Lett.}\ }\textbf {\bibinfo {volume} {131}},\
  \bibinfo {pages} {071001} (\bibinfo {year} {2023})},\ \Eprint
  {https://arxiv.org/abs/2104.02068} {arXiv:2104.02068 [astro-ph.HE]}
  \BibitemShut {NoStop}%
\bibitem [{\citenamefont {Garani}\ and\ \citenamefont
  {Palomares-Ruiz}(2022)}]{Garani:2021feo}%
  \BibitemOpen
  \bibfield  {author} {\bibinfo {author} {\bibfnamefont {R.}~\bibnamefont
  {Garani}}\ and\ \bibinfo {author} {\bibfnamefont {S.}~\bibnamefont
  {Palomares-Ruiz}},\ }\href {https://doi.org/10.1088/1475-7516/2022/05/042}
  {\bibfield  {journal} {\bibinfo  {journal} {JCAP}\ }\textbf {\bibinfo
  {volume} {05}}\bibfield  {number} {\bibinfo  {number} { (05)},\ \bibinfo
  {pages} {042}},\ }\Eprint {https://arxiv.org/abs/2104.12757}
  {arXiv:2104.12757 [hep-ph]} \BibitemShut {NoStop}%
\bibitem [{\citenamefont {Li}\ and\ \citenamefont {Fan}(2022)}]{Li:2022wix}%
  \BibitemOpen
  \bibfield  {author} {\bibinfo {author} {\bibfnamefont {L.}~\bibnamefont
  {Li}}\ and\ \bibinfo {author} {\bibfnamefont {J.}~\bibnamefont {Fan}},\
  }\href {https://doi.org/10.1007/JHEP10(2022)186} {\bibfield  {journal}
  {\bibinfo  {journal} {JHEP}\ }\textbf {\bibinfo {volume} {10}},\ \bibinfo
  {pages} {186}},\ \Eprint {https://arxiv.org/abs/2207.13709} {arXiv:2207.13709
  [hep-ph]} \BibitemShut {NoStop}%
\bibitem [{\citenamefont {French}\ and\ \citenamefont
  {Sher}(2022)}]{French:2022ccb}%
  \BibitemOpen
  \bibfield  {author} {\bibinfo {author} {\bibfnamefont {G.~M.}\ \bibnamefont
  {French}}\ and\ \bibinfo {author} {\bibfnamefont {M.}~\bibnamefont {Sher}},\
  }\href {https://doi.org/10.1103/PhysRevD.106.115037} {\bibfield  {journal}
  {\bibinfo  {journal} {Phys. Rev. D}\ }\textbf {\bibinfo {volume} {106}},\
  \bibinfo {pages} {115037} (\bibinfo {year} {2022})},\ \Eprint
  {https://arxiv.org/abs/2210.04761} {arXiv:2210.04761 [hep-ph]} \BibitemShut
  {NoStop}%
\bibitem [{\citenamefont {Blanco}\ and\ \citenamefont
  {Leane}(2023)}]{Blanco:2023qgi}%
  \BibitemOpen
  \bibfield  {author} {\bibinfo {author} {\bibfnamefont {C.}~\bibnamefont
  {Blanco}}\ and\ \bibinfo {author} {\bibfnamefont {R.~K.}\ \bibnamefont
  {Leane}},\ }\href@noop {} {\  (\bibinfo {year} {2023})},\ \Eprint
  {https://arxiv.org/abs/2312.06758} {arXiv:2312.06758 [hep-ph]} \BibitemShut
  {NoStop}%
\bibitem [{\citenamefont {{Rhodes}}(2019)}]{2019ascl.soft07024R}%
  \BibitemOpen
  \bibfield  {author} {\bibinfo {author} {\bibfnamefont {B.}~\bibnamefont
  {{Rhodes}}},\ }\href@noop {} {\bibinfo {title} {{Skyfield: High precision
  research-grade positions for planets and Earth satellites generator}}},\
  \bibinfo {howpublished} {Astrophysics Source Code Library, record
  ascl:1907.024} (\bibinfo {year} {2019}),\ \Eprint
  {https://arxiv.org/abs/1907.024} {ascl:1907.024} \BibitemShut {NoStop}%
\bibitem [{\citenamefont {Appel}\ \emph {et~al.}(2022)\citenamefont {Appel}
  \emph {et~al.}}]{BOREXINO:2022abl}%
  \BibitemOpen
  \bibfield  {author} {\bibinfo {author} {\bibfnamefont {S.}~\bibnamefont
  {Appel}} \emph {et~al.} (\bibinfo {collaboration} {BOREXINO}),\ }\href
  {https://doi.org/10.1103/PhysRevLett.129.252701} {\bibfield  {journal}
  {\bibinfo  {journal} {Phys. Rev. Lett.}\ }\textbf {\bibinfo {volume} {129}},\
  \bibinfo {pages} {252701} (\bibinfo {year} {2022})},\ \Eprint
  {https://arxiv.org/abs/2205.15975} {arXiv:2205.15975 [hep-ex]} \BibitemShut
  {NoStop}%
\bibitem [{\citenamefont {Gonzalez-Garcia}\ \emph {et~al.}(2023)\citenamefont
  {Gonzalez-Garcia}, \citenamefont {Maltoni}, \citenamefont {Pinheiro},\ and\
  \citenamefont {Serenelli}}]{Gonzalez-Garcia:2023kva}%
  \BibitemOpen
  \bibfield  {author} {\bibinfo {author} {\bibfnamefont {M.~C.}\ \bibnamefont
  {Gonzalez-Garcia}}, \bibinfo {author} {\bibfnamefont {M.}~\bibnamefont
  {Maltoni}}, \bibinfo {author} {\bibfnamefont {J.~a.~P.}\ \bibnamefont
  {Pinheiro}},\ and\ \bibinfo {author} {\bibfnamefont {A.~M.}\ \bibnamefont
  {Serenelli}},\ }\Eprint {https://arxiv.org/abs/2311.16226} {arXiv:2311.16226
  [hep-ph]}  (\bibinfo {year} {2023})\BibitemShut {NoStop}%
\bibitem [{\citenamefont {Esteban}\ \emph {et~al.}(2020)\citenamefont
  {Esteban}, \citenamefont {Gonzalez-Garcia}, \citenamefont {Maltoni},
  \citenamefont {Schwetz},\ and\ \citenamefont {Zhou}}]{Esteban:2020cvm}%
  \BibitemOpen
  \bibfield  {author} {\bibinfo {author} {\bibfnamefont {I.}~\bibnamefont
  {Esteban}}, \bibinfo {author} {\bibfnamefont {M.~C.}\ \bibnamefont
  {Gonzalez-Garcia}}, \bibinfo {author} {\bibfnamefont {M.}~\bibnamefont
  {Maltoni}}, \bibinfo {author} {\bibfnamefont {T.}~\bibnamefont {Schwetz}},\
  and\ \bibinfo {author} {\bibfnamefont {A.}~\bibnamefont {Zhou}},\ }\href
  {https://doi.org/10.1007/JHEP09(2020)178} {\bibfield  {journal} {\bibinfo
  {journal} {JHEP}\ }\textbf {\bibinfo {volume} {09}},\ \bibinfo {pages}
  {178}},\ \Eprint {https://arxiv.org/abs/2007.14792} {arXiv:2007.14792
  [hep-ph]} \BibitemShut {NoStop}%
\bibitem [{\citenamefont {{Torrado}}\ and\ \citenamefont
  {{Lewis}}(2019)}]{2019ascl.soft10019T}%
  \BibitemOpen
  \bibfield  {author} {\bibinfo {author} {\bibfnamefont {J.}~\bibnamefont
  {{Torrado}}}\ and\ \bibinfo {author} {\bibfnamefont {A.}~\bibnamefont
  {{Lewis}}},\ }\href@noop {} {\bibinfo {title} {{Cobaya: Bayesian analysis in
  cosmology}}},\ \bibinfo {howpublished} {Astrophysics Source Code Library,
  record ascl:1910.019} (\bibinfo {year} {2019}),\ \Eprint
  {https://arxiv.org/abs/1910.019} {ascl:1910.019} \BibitemShut {NoStop}%
\bibitem [{\citenamefont {Torrado}\ and\ \citenamefont
  {Lewis}(2021)}]{Torrado:2020dgo}%
  \BibitemOpen
  \bibfield  {author} {\bibinfo {author} {\bibfnamefont {J.}~\bibnamefont
  {Torrado}}\ and\ \bibinfo {author} {\bibfnamefont {A.}~\bibnamefont
  {Lewis}},\ }\href {https://doi.org/10.1088/1475-7516/2021/05/057} {\bibfield
  {journal} {\bibinfo  {journal} {JCAP}\ }\textbf {\bibinfo {volume} {05}},\
  \bibinfo {pages} {057}},\ \Eprint {https://arxiv.org/abs/2005.05290}
  {arXiv:2005.05290 [astro-ph.IM]} \BibitemShut {NoStop}%
\bibitem [{\citenamefont {Basilico}\ \emph {et~al.}(2023)\citenamefont
  {Basilico} \emph {et~al.}}]{BOREXINO:2023ygs}%
  \BibitemOpen
  \bibfield  {author} {\bibinfo {author} {\bibfnamefont {D.}~\bibnamefont
  {Basilico}} \emph {et~al.} (\bibinfo {collaboration} {BOREXINO}),\ }\href
  {https://doi.org/10.1103/PhysRevD.108.102005} {\bibfield  {journal} {\bibinfo
   {journal} {Phys. Rev. D}\ }\textbf {\bibinfo {volume} {108}},\ \bibinfo
  {pages} {102005} (\bibinfo {year} {2023})},\ \Eprint
  {https://arxiv.org/abs/2307.14636} {arXiv:2307.14636 [hep-ex]} \BibitemShut
  {NoStop}%
\bibitem [{\citenamefont {Aharmim}\ \emph
  {et~al.}(2005{\natexlab{a}})\citenamefont {Aharmim} \emph
  {et~al.}}]{SNO:2005ftm}%
  \BibitemOpen
  \bibfield  {author} {\bibinfo {author} {\bibfnamefont {B.}~\bibnamefont
  {Aharmim}} \emph {et~al.} (\bibinfo {collaboration} {SNO}),\ }\href
  {https://doi.org/10.1103/PhysRevD.72.052010} {\bibfield  {journal} {\bibinfo
  {journal} {Phys. Rev. D}\ }\textbf {\bibinfo {volume} {72}},\ \bibinfo
  {pages} {052010} (\bibinfo {year} {2005}{\natexlab{a}})},\ \Eprint
  {https://arxiv.org/abs/hep-ex/0507079} {arXiv:hep-ex/0507079} \BibitemShut
  {NoStop}%
\bibitem [{\citenamefont {Aharmim}\ \emph
  {et~al.}(2005{\natexlab{b}})\citenamefont {Aharmim} \emph
  {et~al.}}]{SNO:2005oxr}%
  \BibitemOpen
  \bibfield  {author} {\bibinfo {author} {\bibfnamefont {B.}~\bibnamefont
  {Aharmim}} \emph {et~al.} (\bibinfo {collaboration} {SNO}),\ }\href
  {https://doi.org/10.1103/PhysRevC.72.055502} {\bibfield  {journal} {\bibinfo
  {journal} {Phys. Rev. C}\ }\textbf {\bibinfo {volume} {72}},\ \bibinfo
  {pages} {055502} (\bibinfo {year} {2005}{\natexlab{b}})},\ \Eprint
  {https://arxiv.org/abs/nucl-ex/0502021} {arXiv:nucl-ex/0502021} \BibitemShut
  {NoStop}%
\bibitem [{\citenamefont {Li}\ \emph {et~al.}(2023)\citenamefont {Li},
  \citenamefont {Liu},\ and\ \citenamefont {Xue}}]{Li:2022idr}%
  \BibitemOpen
  \bibfield  {author} {\bibinfo {author} {\bibfnamefont {Y.}~\bibnamefont
  {Li}}, \bibinfo {author} {\bibfnamefont {Z.}~\bibnamefont {Liu}},\ and\
  \bibinfo {author} {\bibfnamefont {Y.}~\bibnamefont {Xue}},\ }\href
  {https://doi.org/10.1088/1475-7516/2023/05/060} {\bibfield  {journal}
  {\bibinfo  {journal} {JCAP}\ }\textbf {\bibinfo {volume} {05}},\ \bibinfo
  {pages} {060}},\ \Eprint {https://arxiv.org/abs/2209.04387} {arXiv:2209.04387
  [hep-ph]} \BibitemShut {NoStop}%
\bibitem [{\citenamefont {Ema}\ \emph {et~al.}(2024)\citenamefont {Ema},
  \citenamefont {Pospelov},\ and\ \citenamefont {Ray}}]{Ema:2024oce}%
  \BibitemOpen
  \bibfield  {author} {\bibinfo {author} {\bibfnamefont {Y.}~\bibnamefont
  {Ema}}, \bibinfo {author} {\bibfnamefont {M.}~\bibnamefont {Pospelov}},\ and\
  \bibinfo {author} {\bibfnamefont {A.}~\bibnamefont {Ray}},\ }\href@noop {} {\
   (\bibinfo {year} {2024})},\ \Eprint {https://arxiv.org/abs/2402.03431}
  {arXiv:2402.03431 [hep-ph]} \BibitemShut {NoStop}%
\bibitem [{\citenamefont {Roy}\ \emph {et~al.}(2024)\citenamefont {Roy},
  \citenamefont {Dasgupta},\ and\ \citenamefont {Guchait}}]{Roy:2024ear}%
  \BibitemOpen
  \bibfield  {author} {\bibinfo {author} {\bibfnamefont {A.}~\bibnamefont
  {Roy}}, \bibinfo {author} {\bibfnamefont {B.}~\bibnamefont {Dasgupta}},\ and\
  \bibinfo {author} {\bibfnamefont {M.}~\bibnamefont {Guchait}},\ }\href@noop
  {} {\  (\bibinfo {year} {2024})},\ \Eprint {https://arxiv.org/abs/2402.17265}
  {arXiv:2402.17265 [hep-ph]} \BibitemShut {NoStop}%
\bibitem [{\citenamefont {Alves}\ \emph {et~al.}(2023)\citenamefont {Alves}
  \emph {et~al.}}]{Alves:2023ree}%
  \BibitemOpen
  \bibfield  {author} {\bibinfo {author} {\bibfnamefont {D.~S.~M.}\
  \bibnamefont {Alves}} \emph {et~al.},\ }\href
  {https://doi.org/10.1140/epjc/s10052-023-11271-x} {\bibfield  {journal}
  {\bibinfo  {journal} {Eur. Phys. J. C}\ }\textbf {\bibinfo {volume} {83}},\
  \bibinfo {pages} {230} (\bibinfo {year} {2023})}\BibitemShut {NoStop}%
\bibitem [{\citenamefont {Chang}\ \emph {et~al.}(2021)\citenamefont {Chang},
  \citenamefont {Essig},\ and\ \citenamefont {Reinert}}]{Chang:2019xva}%
  \BibitemOpen
  \bibfield  {author} {\bibinfo {author} {\bibfnamefont {J.~H.}\ \bibnamefont
  {Chang}}, \bibinfo {author} {\bibfnamefont {R.}~\bibnamefont {Essig}},\ and\
  \bibinfo {author} {\bibfnamefont {A.}~\bibnamefont {Reinert}},\ }\href
  {https://doi.org/10.1007/JHEP03(2021)141} {\bibfield  {journal} {\bibinfo
  {journal} {JHEP}\ }\textbf {\bibinfo {volume} {03}},\ \bibinfo {pages}
  {141}},\ \Eprint {https://arxiv.org/abs/1911.03389} {arXiv:1911.03389
  [hep-ph]} \BibitemShut {NoStop}%
\bibitem [{\citenamefont {Gninenko}\ and\ \citenamefont
  {Krasnikov}(1998)}]{Gninenko:1998pm}%
  \BibitemOpen
  \bibfield  {author} {\bibinfo {author} {\bibfnamefont {S.~N.}\ \bibnamefont
  {Gninenko}}\ and\ \bibinfo {author} {\bibfnamefont {N.~V.}\ \bibnamefont
  {Krasnikov}},\ }\href {https://doi.org/10.1016/S0370-2693(98)00358-X}
  {\bibfield  {journal} {\bibinfo  {journal} {Phys. Lett. B}\ }\textbf
  {\bibinfo {volume} {427}},\ \bibinfo {pages} {307} (\bibinfo {year}
  {1998})},\ \Eprint {https://arxiv.org/abs/hep-ph/9802375}
  {arXiv:hep-ph/9802375} \BibitemShut {NoStop}%
\bibitem [{\citenamefont {Altegoer}\ \emph {et~al.}(1998)\citenamefont
  {Altegoer} \emph {et~al.}}]{NOMAD:1998pxi}%
  \BibitemOpen
  \bibfield  {author} {\bibinfo {author} {\bibfnamefont {J.}~\bibnamefont
  {Altegoer}} \emph {et~al.} (\bibinfo {collaboration} {NOMAD}),\ }\href
  {https://doi.org/10.1016/S0370-2693(98)00402-X} {\bibfield  {journal}
  {\bibinfo  {journal} {Phys. Lett. B}\ }\textbf {\bibinfo {volume} {428}},\
  \bibinfo {pages} {197} (\bibinfo {year} {1998})},\ \Eprint
  {https://arxiv.org/abs/hep-ex/9804003} {arXiv:hep-ex/9804003} \BibitemShut
  {NoStop}%
\bibitem [{\citenamefont {Aprile}\ \emph {et~al.}(2022)\citenamefont {Aprile}
  \emph {et~al.}}]{XENON:2022ltv}%
  \BibitemOpen
  \bibfield  {author} {\bibinfo {author} {\bibfnamefont {E.}~\bibnamefont
  {Aprile}} \emph {et~al.} (\bibinfo {collaboration} {XENON}),\ }\href
  {https://doi.org/10.1103/PhysRevLett.129.161805} {\bibfield  {journal}
  {\bibinfo  {journal} {Phys. Rev. Lett.}\ }\textbf {\bibinfo {volume} {129}},\
  \bibinfo {pages} {161805} (\bibinfo {year} {2022})},\ \Eprint
  {https://arxiv.org/abs/2207.11330} {arXiv:2207.11330 [hep-ex]} \BibitemShut
  {NoStop}%
\bibitem [{\citenamefont {Ansarifard}\ and\ \citenamefont
  {Farzan}(2023)}]{Ansarifard:2022kvy}%
  \BibitemOpen
  \bibfield  {author} {\bibinfo {author} {\bibfnamefont {S.}~\bibnamefont
  {Ansarifard}}\ and\ \bibinfo {author} {\bibfnamefont {Y.}~\bibnamefont
  {Farzan}},\ }\href {https://doi.org/10.1103/PhysRevD.107.075029} {\bibfield
  {journal} {\bibinfo  {journal} {Phys. Rev. D}\ }\textbf {\bibinfo {volume}
  {107}},\ \bibinfo {pages} {075029} (\bibinfo {year} {2023})},\ \Eprint
  {https://arxiv.org/abs/2211.09105} {arXiv:2211.09105 [hep-ph]} \BibitemShut
  {NoStop}%
\bibitem [{\citenamefont {Anamiati}\ \emph {et~al.}(2018)\citenamefont
  {Anamiati}, \citenamefont {Fonseca},\ and\ \citenamefont
  {Hirsch}}]{Anamiati:2017rxw}%
  \BibitemOpen
  \bibfield  {author} {\bibinfo {author} {\bibfnamefont {G.}~\bibnamefont
  {Anamiati}}, \bibinfo {author} {\bibfnamefont {R.~M.}\ \bibnamefont
  {Fonseca}},\ and\ \bibinfo {author} {\bibfnamefont {M.}~\bibnamefont
  {Hirsch}},\ }\href {https://doi.org/10.1103/PhysRevD.97.095008} {\bibfield
  {journal} {\bibinfo  {journal} {Phys. Rev. D}\ }\textbf {\bibinfo {volume}
  {97}},\ \bibinfo {pages} {095008} (\bibinfo {year} {2018})},\ \Eprint
  {https://arxiv.org/abs/1710.06249} {arXiv:1710.06249 [hep-ph]} \BibitemShut
  {NoStop}%
\bibitem [{\citenamefont {Abe}\ \emph {et~al.}(2024{\natexlab{a}})\citenamefont
  {Abe} \emph {et~al.}}]{Super-Kamiokande:2023jbt}%
  \BibitemOpen
  \bibfield  {author} {\bibinfo {author} {\bibfnamefont {K.}~\bibnamefont
  {Abe}} \emph {et~al.} (\bibinfo {collaboration} {Super-Kamiokande}),\ }\href
  {https://doi.org/10.1103/PhysRevD.109.092001} {\bibfield  {journal} {\bibinfo
   {journal} {Phys. Rev. D}\ }\textbf {\bibinfo {volume} {109}},\ \bibinfo
  {pages} {092001} (\bibinfo {year} {2024}{\natexlab{a}})},\ \Eprint
  {https://arxiv.org/abs/2312.12907} {arXiv:2312.12907 [hep-ex]} \BibitemShut
  {NoStop}%
\bibitem [{\citenamefont {Abe}\ \emph {et~al.}(2024{\natexlab{b}})\citenamefont
  {Abe} \emph {et~al.}}]{Super-Kamiokande:2023yqq}%
  \BibitemOpen
  \bibfield  {author} {\bibinfo {author} {\bibfnamefont {K.}~\bibnamefont
  {Abe}} \emph {et~al.} (\bibinfo {collaboration} {Super-Kamiokande}),\ }\href
  {https://doi.org/10.1103/PhysRevLett.132.241803} {\bibfield  {journal}
  {\bibinfo  {journal} {Phys. Rev. Lett.}\ }\textbf {\bibinfo {volume} {132}},\
  \bibinfo {pages} {241803} (\bibinfo {year} {2024}{\natexlab{b}})},\ \Eprint
  {https://arxiv.org/abs/2311.01159} {arXiv:2311.01159 [hep-ex]} \BibitemShut
  {NoStop}%
\bibitem [{\citenamefont {Buckley}\ \emph {et~al.}(2015)\citenamefont
  {Buckley}, \citenamefont {Ferrando}, \citenamefont {Lloyd}, \citenamefont
  {Nordstr\"om}, \citenamefont {Page}, \citenamefont {R\"ufenacht},
  \citenamefont {Sch\"onherr},\ and\ \citenamefont
  {Watt}}]{buckley2015lhapdf6}%
  \BibitemOpen
  \bibfield  {author} {\bibinfo {author} {\bibfnamefont {A.}~\bibnamefont
  {Buckley}}, \bibinfo {author} {\bibfnamefont {J.}~\bibnamefont {Ferrando}},
  \bibinfo {author} {\bibfnamefont {S.}~\bibnamefont {Lloyd}}, \bibinfo
  {author} {\bibfnamefont {K.}~\bibnamefont {Nordstr\"om}}, \bibinfo {author}
  {\bibfnamefont {B.}~\bibnamefont {Page}}, \bibinfo {author} {\bibfnamefont
  {M.}~\bibnamefont {R\"ufenacht}}, \bibinfo {author} {\bibfnamefont
  {M.}~\bibnamefont {Sch\"onherr}},\ and\ \bibinfo {author} {\bibfnamefont
  {G.}~\bibnamefont {Watt}},\ }\href
  {https://doi.org/10.1140/epjc/s10052-015-3318-8} {\bibfield  {journal}
  {\bibinfo  {journal} {Eur. Phys. J. C}\ }\textbf {\bibinfo {volume} {75}},\
  \bibinfo {pages} {132} (\bibinfo {year} {2015})},\ \Eprint
  {https://arxiv.org/abs/1412.7420} {arXiv:1412.7420 [hep-ph]} \BibitemShut
  {NoStop}%
\bibitem [{\citenamefont {Lewis}(2019)}]{Lewis:2019xzd}%
  \BibitemOpen
  \bibfield  {author} {\bibinfo {author} {\bibfnamefont {A.}~\bibnamefont
  {Lewis}},\ }\href {https://getdist.readthedocs.io} {\bibinfo {title}
  {{GetDist: a Python package for analysing Monte Carlo samples}}} (\bibinfo
  {year} {2019}),\ \Eprint {https://arxiv.org/abs/1910.13970} {arXiv:1910.13970
  [astro-ph.IM]} \BibitemShut {NoStop}%
\bibitem [{\citenamefont {Bringmann}\ and\ \citenamefont
  {Pospelov}(2019)}]{Bringmann:2018cvk}%
  \BibitemOpen
  \bibfield  {author} {\bibinfo {author} {\bibfnamefont {T.}~\bibnamefont
  {Bringmann}}\ and\ \bibinfo {author} {\bibfnamefont {M.}~\bibnamefont
  {Pospelov}},\ }\href {https://doi.org/10.1103/PhysRevLett.122.171801}
  {\bibfield  {journal} {\bibinfo  {journal} {Phys. Rev. Lett.}\ }\textbf
  {\bibinfo {volume} {122}},\ \bibinfo {pages} {171801} (\bibinfo {year}
  {2019})},\ \Eprint {https://arxiv.org/abs/1810.10543} {arXiv:1810.10543
  [hep-ph]} \BibitemShut {NoStop}%
\bibitem [{\citenamefont {Cappiello}\ and\ \citenamefont
  {Beacom}(2019)}]{Cappiello:2019qsw}%
  \BibitemOpen
  \bibfield  {author} {\bibinfo {author} {\bibfnamefont {C.~V.}\ \bibnamefont
  {Cappiello}}\ and\ \bibinfo {author} {\bibfnamefont {J.~F.}\ \bibnamefont
  {Beacom}},\ }\href {https://doi.org/10.1103/PhysRevD.104.069901} {\bibfield
  {journal} {\bibinfo  {journal} {Phys. Rev. D}\ }\textbf {\bibinfo {volume}
  {100}},\ \bibinfo {pages} {103011} (\bibinfo {year} {2019})},\ \bibinfo
  {note} {[Erratum: Phys.Rev.D 104, 069901 (2021)]},\ \Eprint
  {https://arxiv.org/abs/1906.11283} {arXiv:1906.11283 [hep-ph]} \BibitemShut
  {NoStop}%
\bibitem [{\citenamefont {Wang}\ \emph {et~al.}(2023)\citenamefont {Wang},
  \citenamefont {Wu}, \citenamefont {Yang},\ and\ \citenamefont
  {Zhu}}]{Wang:2021nbf}%
  \BibitemOpen
  \bibfield  {author} {\bibinfo {author} {\bibfnamefont {W.}~\bibnamefont
  {Wang}}, \bibinfo {author} {\bibfnamefont {L.}~\bibnamefont {Wu}}, \bibinfo
  {author} {\bibfnamefont {W.-N.}\ \bibnamefont {Yang}},\ and\ \bibinfo
  {author} {\bibfnamefont {B.}~\bibnamefont {Zhu}},\ }\href
  {https://doi.org/10.1103/PhysRevD.107.073002} {\bibfield  {journal} {\bibinfo
   {journal} {Phys. Rev. D}\ }\textbf {\bibinfo {volume} {107}},\ \bibinfo
  {pages} {073002} (\bibinfo {year} {2023})},\ \Eprint
  {https://arxiv.org/abs/2111.04000} {arXiv:2111.04000 [hep-ph]} \BibitemShut
  {NoStop}%
\bibitem [{\citenamefont {Cui}\ \emph {et~al.}(2022)\citenamefont {Cui} \emph
  {et~al.}}]{PandaX-II:2021kai}%
  \BibitemOpen
  \bibfield  {author} {\bibinfo {author} {\bibfnamefont {X.}~\bibnamefont
  {Cui}} \emph {et~al.} (\bibinfo {collaboration} {PandaX-II}),\ }\href
  {https://doi.org/10.1103/PhysRevLett.128.171801} {\bibfield  {journal}
  {\bibinfo  {journal} {Phys. Rev. Lett.}\ }\textbf {\bibinfo {volume} {128}},\
  \bibinfo {pages} {171801} (\bibinfo {year} {2022})},\ \Eprint
  {https://arxiv.org/abs/2112.08957} {arXiv:2112.08957 [hep-ex]} \BibitemShut
  {NoStop}%
\bibitem [{\citenamefont {Xu}\ \emph {et~al.}(2022)\citenamefont {Xu} \emph
  {et~al.}}]{CDEX:2022fig}%
  \BibitemOpen
  \bibfield  {author} {\bibinfo {author} {\bibfnamefont {R.}~\bibnamefont {Xu}}
  \emph {et~al.} (\bibinfo {collaboration} {CDEX}),\ }\href
  {https://doi.org/10.1103/PhysRevD.106.052008} {\bibfield  {journal} {\bibinfo
   {journal} {Phys. Rev. D}\ }\textbf {\bibinfo {volume} {106}},\ \bibinfo
  {pages} {052008} (\bibinfo {year} {2022})},\ \Eprint
  {https://arxiv.org/abs/2201.01704} {arXiv:2201.01704 [hep-ex]} \BibitemShut
  {NoStop}%
\end{thebibliography}%

\end{document}